# Anomalous Frequency Trends in MoS$_2$ Thin Films Attributed to Surface Effects

Xin Luo[1], Yanyuan Zhao[2], Jun Zhang[2], Qihua Xiong[2,3,*], Su Ying Quek[1,†,*]

[1]Institute of High Performance Computing, 1 Fusionopolis Way, #16-16 Connexis, Singapore 138632

[2]Division of Physics and Applied Physics, School of Physical and Mathematical Sciences, Nanyang Technological University, 21 Nanyang Link, Singapore 637371

[3]Division of Microelectronics, School of Electrical and Electronic Engineering, Nanyang Technological University, Singapore, 639798

*: To whom correspondence should be addressed. Email address: qihua@ntu.edu.sg (experiment), phyqsy@nus.edu.sg (theory)
†: Present address: Department of Physics, National University of Singapore, S12-M01, 2 Science Drive 3, Singapore 117551

## Abstract

The layered dichalcogenide MoS$_2$ has many unique physical properties in low dimensions. Recent experimental Raman spectroscopies report an anomalous blue shift of the in-plane $E_{2g}^1$ mode with decreasing thickness, a trend that is not understood. Here, we combine experimental Raman scattering and theoretical studies to clarify and explain this trend. Special attention is given to understanding the surface effect on Raman frequencies by using a force constants model based on first-principles calculations. Surface effects refer to the larger Mo-S force constants at the surface of thin film MoS$_2$, which results from a loss of neighbours in adjacent MoS$_2$ layers. Without surface effects, the frequencies of both out-of-plane $A_{1g}$ and in-plane $E_{2g}^1$ modes decrease with decreasing thickness. However, the $E_{2g}^1$ mode blue shifts while the $A_{1g}$ mode red shifts once the surface effect is **included, in agreement with the experiment. Our results show that competition between the thickness effect and the surface effect determines the mechanical properties of two-dimensional MoS$_2$, which we believe applies to other layered materials.**

KEYWORDS: Raman Spectrum, Density Functional Theory, 2D Transition-metal Dichalcogenide, Force Constants model, Surface Effect.

PACS number(s): 63.20.dk, 68.35.Gy, 71.15.Mb

## I. INTRODUCTION

MoS$_2$, a prototypical transition metal dichalcogenide semiconductor, comprises weakly interacting trilayers (TL, with three atomic layers in one TL) in its most common 2H structure (see Fig. 1a-b).[1,2] Due to the weak van der Waals (vdW) interactions between TLs, thin films can be easily prepared using mechanical exfoliation. MoS$_2$ thin films and nanoparticles have attracted great interest.[3-9] In these low-dimensional structures, the environment of atoms at the surface is different from that in the interior or in the bulk, leading to different physical and chemical properties. For instance, the increased surface-to-volume ratio in MoS$_2$ nanoclusters[10] and surfaces[11] can greatly improve the catalytic activity for the hydrogen evolution reaction. Another recent paper demonstrated that there is an indirect-to-direct band gap transition when decreasing the film thickness to a single TL (1 TL), together with a significant enhancement of photoluminescence quantum yield in 1 TL.[12] Moreover, recent experiments[13,14] suggest the possibility of valley-based electronic and optoelectronic applications in MoS$_2$ low-dimensional systems. These unique properties of thin film MoS$_2$ are expected to be sensitive to its thickness.

Raman spectroscopy is widely used in the characterization of low-dimensional thin film systems, because the phonon frequencies are sensitive to the thickness of these thin films.[15-19]

Recent Raman experiments on few-TL MoS$_2$ reveal several consistent trends for the low-frequency phonon modes as a function of thickness, attributed to interlayer shear and breathing modes.[17,18,20,21] However, within the high-frequency range, which can be easier to probe, there are contradictory experimental reports regarding how the frequency of the in-plane $E_{2g}^1$ mode changes from bulk to two-dimensional (2D). Specifically, Lee *et al.* found that while the out-of-plane $A_{1g}$ mode (as will be shown in Fig. 5(a)) red shifts with decreasing thickness, the in-plane $E_{2g}^1$ mode (as will be shown in Fig. 5(d)) blue shifts.[16] The blue shift of the $E_{2g}^1$ mode was thought to be anomalous,[16] because this mode corresponds to an in-plane motion with atoms displacing out of phase, which should intuitively result in lower frequencies as the thickness (and therefore "friction") decreases due to a smaller restoring force. This result was subsequently reproduced by Li *et al.*[22] and Zeng *et al.*[20] using incident excitation lasers with different wavelengths. However, another experimental group recently reported a different trend in the $E_{2g}^1$ modes, where both the $A_{1g}$ and $E_{2g}^1$ modes red shift when the sample goes from three dimensional (3D) to 2D.[23]

On the theoretical front, it was found that density functional theory (DFT) calculations with the local density approximation (LDA) for the exchange-correlation functional could predict the ultralow-frequency phonon modes in few-TL MoS$_2$, in good agreement with the experimental observations.[17,20] At high-frequency, DFT calculations with different exchange-correlation functionals give different results; for example, both the $A_{1g}$ and the $E_{2g}^1$ modes blue shift from 3D to 2D in the generalized gradient approximation with dispersion correction,[24] while the $A_{1g}$ mode red shifts and the $E_{2g}^1$ mode blue shifts using LDA.[24,25]

While there are more experiments pointing toward an anomalous blue shift for the $E_{2g}^1$ modes, it is not clear why this occurs. Similar anomalous Raman evolution trends of the $E_{2g}^1$ mode are found in the other transition metal dichalcogenide (TMD) materials such as $MoSe_2$,[26,27] $WS_2$,[28] and $WSe_2$[17,29]. Experimental and theoretical papers have discussed that the anomalous trend reflects the presence of additional interactions, and possible explanations for this trend are stacking-induced structural changes or long-range Coulomb interactions.[16,25] However, the exact mechanism is still not understood.[21,30,31] Given the intense interest in 2D $MoS_2$ for diverse applications, and potential applications in nanoscale electromechanical devices, a systematic study on the lattice vibrational modes of 2D $MoS_2$ has become increasingly important. What physical and chemical mechanisms influence the frequency trends in these 2D materials? How does (1) moving from bulk to thin films and (2) changing the thickness of thin films affect these mechanisms?

In this paper, we present both DFT and experimental Raman scattering studies on lattice vibrations of few-TL and bulk $MoS_2$. Based on the DFT calculation, a force constants model is developed to gain physical insights into the mechanisms that determine the frequency trends. The theoretical controversy over the frequency trends arises from the uncertainty over which exchange-correlation functional in DFT is most appropriate to describe and predict the frequencies. By benchmarking our calculations against the well-established high-frequency phonon modes in the bulk, we conclude that DFT-LDA is the most appropriate, a finding that is consistent with our previous conclusion[17] for the ultralow frequencies in few-TL $MoS_2$. Based on our theoretical and experimental studies, we conclude that the $E_{2g}^1$ mode blue shifts while the $A_{1g}$ mode red shifts as the number of TLs in thin film $MoS_2$ decreases. Our analysis indicates that although MoS$_2$ is a layered material with only weak vdW interactions between layers, these weak vdW interactions have subtle but evident effects in the observed frequency trends in few-TL MoS$_2$. Specifically, S atoms at the surface lose their neighbors in adjacent TLs and this results in stronger Mo-S force constants at the surface. Our first-principles-based force constants model indicates that the frequencies of the $E_{2g}^1$ and $A_{1g}$ modes are determined by a competition between the effect of increased force constants between Mo and S atoms at the surface, and the effect of decreasing thickness (i.e. decreasing "friction" between layers). The effect of increased surface force constants is larger than the effect of decreasing thickness for the $E_{2g}^1$ mode, explaining its anomalous blue shift.

## II. METHODS

First-principles calculations are performed using DFT as implemented in the plane-wave pseudopotential code Quantum Open-Source Package for Research in Electronic Structure Simulation and Optimization (ESPRESSO).[32] Based on our benchmark studies, we use the LDA[33] to the exchange-correlation functional for most of our results. The benchmark studies are also performed using a semiempirical vdW correction based on Grimme's scheme (Perdew-Burke-Ernzerhof functional including dispersion correction, or PBE-D2)[34] and the van der Waals density functional (vdW-DF)[35] with Cooper's exchange[36]. To get converged results, a plane-wave kinetic energy cutoff of 65 Ry is used for the LDA wave functions, and a higher cutoff of 80 Ry is used for PBE-D2 and vdW-DF wave functions. For the few-TL systems, neighboring slabs are separated by a vacuum thickness of 16 Å (this value has been tested for convergence of phonon frequencies). The Monkhorst-Pack $k$-point meshes of 17×17×5 and 17×17×1 are used to sample the Brillouin Zones for bulk and thin film systems, respectively. In the self-consistent calculation, the convergence threshold for energy is set to $10^{-9}$ eV. All atomic coordinates and lattice constants are fully relaxed until the absolute value of force acting on each atom is less than 0.003 eV/Å. During the structure optimization, the hexagonal symmetry is imposed on the bulk and few-TL systems. With the optimized structures and self-consistent wave functions, the phonon spectrum and Raman intensities are calculated within density-functional perturbation theory (DFPT) as introduced by Lazzeri and Mauri.[37]

We calculate the Raman intensities of the phonon modes within DFPT. Within the Placzek approximation[38], the nonresonant Raman intensity ($\omega_{Laser} \gg \omega_p$) is found to be proportional to $|e_i \cdot \tilde{R} \cdot e_s|^2$, where $e_i$ is the polarization vector of the incident light and $e_s$ is that of the scattered light. $\tilde{R}_{ij}$ is the second-rank polarizability tensor, which is obtained from the second-order derivative of the electronic density matrix with respect to a uniform electric field.[37] A given phonon mode can be observed by Raman scattering spectroscopy only when $|e_i \cdot \tilde{R} \cdot e_s|^2$ does not equal to zero. In experiments, both in-plane and out-of-plane vibrational modes can be detected when the polarizations of incident and scattering laser light are in the $x$ direction parallel to the TL plane. Therefore, we compute the Raman intensities under this polarization configuration ($I_{xx}$), as will be presented in Table III.

The experimental samples of 1 TL and few-TL $MoS_2$ are obtained from bulk crystals by scotch tape-based mechanical exfoliation and are then deposited onto freshly cleaned Si substrates with a 90 nm thick $SiO_2$. The sample thicknesses are determined by using atomic force microscopy. Raman scattering spectroscopy measurements are carried out at room temperature using a micro-Raman spectrometer (Horiba-JY T64000) equipped with a liquid nitrogen-cooled charge-coupled device. The measurements were conducted in a backscattering configuration excited with a solid state green laser (λ = 532 nm). More experimental details are available in Ref.17.

## III. RESULTS AND DISCUSSION

### A. Choice of Exchange-Correlation Functional

As a starting point for choosing the appropriate DFT exchange-correlation functional, we report the optimized structural parameters for bulk and 1 TL MoS$_2$ (Table I). We define the bulk in-plane lattice parameter to be $a$ and the out-of-plane lattice parameter to be $c$. Compared with the experimental lattice parameters of bulk MoS$_2$ ($a$=3.160 Å, $c$=12.294 Å),[39] the LDA calculation gives a close in-plane lattice parameter (error within 0.3% for $a$) but underestimates $c$ by 1.6%. However, the vdW-corrected functionals give a larger deviation on $a$, but a more accurate prediction for $c$. Low-energy electron diffraction experiments have observed[40] that the interplanar spacing between Mo and S atoms at the surface is smaller than that in the bulk. From Table I and calculations on 2- to 9- TL MoS$_2$, we find that both LDA functional and vdW-DF properly capture this effect, in contrast to PBE-D2, where surface Mo-S bond lengths are longer than the interior ones. Comparing in Table 2 the predicted $\Gamma$-point phonon frequencies with the experiment,[41-43] we find that in general, PBE-D2 underestimates all frequencies, vdW-DF overestimates frequencies in the low-frequency range and underestimates frequencies in the high-frequency range, and LDA gives the best match with the experiment. Importantly, LDA also gives the best estimate for the well-established frequency difference in the ($E_{1u}^2$, $E_{2g}^1$) Davydov doublet (discussed below; Table II). These benchmark studies therefore indicate that although LDA does not explicitly include vdW interactions, it best reproduces the change in force with respect to ionic displacements, *i.e.* the force constants and frequencies. This finding is consistent with other studies on layered compounds,[17,44-47] and from here onward, we focus on the LDA results.

## B. Frequencies in the Bulk

We start by examining the factors that influence the frequency trends in bulk phonons. According to group theory,[48] the $\Gamma$–point phonon modes of bulk MoS$_2$ can be decomposed into the irreducible representation $\Gamma = 2A_{2u} + 2E_{1u} + 2B_{2g} + 2E_{2g} + A_{1g} + E_{1g} + B_{1u} + E_{2u}$. These 12 irreducible representations in bulk MoS$_2$ can be divided into six conjugate pairs, as shown in Fig. 1(c). In each conjugate pair, there is an inversion center between the two TLs, and one phonon mode differs from the other by an inter-TL phase shift of 180°. There are three in-plane conjugate pairs: ($E_{1u}^1$, $E_{2g}^2$), ($E_{1g}$, $E_{2u}$), and ($E_{1u}^2$, $E_{2g}^1$), among which the ($E_{1g}$, $E_{2u}$) and ($E_{1u}^2$, $E_{2g}^1$) conjugate modes have quite similar energy, and are known as Davydov doublets.[42,48] The Davydov splitting is the frequency difference (~ 1-2 cm$^{-1}$) between two modes in a Davydov doublet.

Understanding the subtle Davydov splitting is an important first step to elucidate the nature of inter-TL interactions. It is natural to think that the frequency of $E_{2g}^1$ should be larger than that of $E_{1u}^2$ (Fig. 1(c)), i.e. $\omega(E_{1u}^2) < \omega(E_{2g}^1)$, and $\omega(E_{2u}) < \omega(E_{1g})$, due to the additional bond stretch involving sulfur atoms in adjacent TLs. A simple oscillator model[49,50] considering only inter-TL vdW interactions predicts these trends. However, experiments find instead that $\omega(E_{1u}^2) > \omega(E_{2g}^1)$. Such anomalous Davydov splitting has also been found in other layered materials, such as GaSe, GaS, and As$_2$Se$_3$.[51,52] Our LDA calculations indicate that the in plane force constant between S and Mo atoms in adjacent TLs is ~40% larger in magnitude than that corresponding to vdW interactions between adjacent S atoms. By including interactions up to the second-nearest neighbors, we can construct a force constants model (see APPENDIX) that gives consistent trends with experiments and LDA frequencies (Table II). Therefore, Coulomb interactions between S and Mo atoms in adjacent TLs are important for explaining the subtle Davydov splitting in ($E_{1u}^2$, $E_{2g}^1$), where both Mo and S atoms are displaced.

### C. Frequency Trends in MoS$_2$ Thin Films

Next, we discuss the phonon frequencies in few-TL MoS$_2$. According to group theory,[17] for $N$ odd, $N$TL MoS$_2$ belongs to space group $D_{3h}^1$ ($P\bar{6}m2$) without inversion symmetry, while for $N$ even, $N$TL MoS$_2$ belongs to space group $D_{3d}^3$ ($P\bar{3}m1$) with inversion symmetry. For $N$ odd, the zone center phonon modes have the irreproducible representation $\Gamma = \frac{(3N-1)}{2}(A_1' + A_2'' + E' + E'') + A_2'' + E', N = 1,3,5...$, and for $N$ even, $\Gamma = \frac{3N}{2}(A_{1g} + A_{2u} + E_g + E_u), N = 2,4,6...$. In each $N$TL MoS$_2$, there are $3N$ twofold degenerate in-plane modes and $3N$ out-of-plane modes. Table III shows the LDA-calculated $\Gamma$-point phonon frequencies in 1-4 TLs and bulk MoS$_2$. Interestingly, each mode in the 1 TL case evolves into $N$ modes in $N$TL MoS$_2$ with similar frequencies, as highlighted by the background color in Table III. By examining the displacement vector of the evolved $N$ modes, we find that the atoms within each TL in the $N$TL system are displaced in the same manner as the original mode in 1 TL but that the $N$ modes are distinguished from one another by the different relative phases among TLs. This is just like the Davydov doublets in the bulk limit, where the inter-TL phases are 0° and 180° between any adjacent TLs (the two extreme cases).

How do our results compare with those of the experiment? For the frequency range 360-440 cm$^{-1}$, our Raman experiments uncover two peaks in 1- to 12- TL MoS$_2$ (Fig. 2), which has been reported in other experiments.[16,20,22] Based on their frequencies, they are evolved from the Raman-active in-plane $E_{2g}^1$ and out-of-plane $A_{1g}$ modes in the bulk. Our nonresonant Raman intensity calculations[37,38] also show that the observed modes are those with large Raman intensity and have the same 180° inter-TL phases (as will be shown in Fig. 5(c) and (f)) as the bulk $E_{2g}^1$ and $A_{1g}$ modes (Fig. 1(c)). Although the modes in $N$TL are, strictly speaking, not $E_{2g}^1$ and $A_{1g}$ modes by symmetry (the correct notation is shown in Table III), we use the same notation to be consistent with the literature. The LDA frequencies are also plotted in Fig. 2 and compare reasonably well with those of the experiment. Our calculation also predicts that the phonon mode with a frequency of 387 cm$^{-1}$ in 3-TL MoS$_2$ has quite a large Raman intensity; it is not detected in experiment, which could be because it is only a shoulder in the broader peak at 385 cm$^{-1}$. Importantly, both our experiments and calculations find that the $E_{2g}^1$ mode blue shifts while the $A_{1g}$ mode red shifts as the number of TLs decreases. A quantitative comparison of the experimental trends will be shown in Fig. 5(a) and 5(d). We suggest that the different experimental results from Ramakrishna Matte $et\ al.$[23] arise from the effects of strain,[53] which are relevant for wrinkles and folds in their chemically synthesized MoS$_2$ thin layers, as shown in their transmission electron microscopy images of the samples.

The $A_{1g}$ peak exhibits a smooth broadening from bulk to the 2D case (Fig. 2), while the line width of the $E_{2g}^1$ peak does not show distinct thickness dependence. The broadening of the $A_{1g}$ peak is a direct result of the phonon confinement effect[54,55] along the z axis and indicates a reduced phonon lifetime in 2D. A similar phenomenon has been reported in other layered materials.[56] The strong thickness dependence of the line width of $A_{1g}$ modes is directly related to its out-of-plane vibration manner (as shown in Fig. 1(c)), along which direction the size is largely confined. However, the lateral size is considered not to change much from bulk to 2D, and thus the in-plane vibrations should not be affected in the same way as the out-of-plane vibrations, explaining the small thickness dependence of the line width of the $E_{2g}^1$ peak. (A more detailed and quantitative analysis is provided in the supplemental material.)

### D. Force Constants

How do we explain the observed frequency trends? We start by analyzing the force constants in the different systems, as computed by DFT-LDA (see APPENDIX for sign convention of force constants). We focus on interactions up to the second-nearest neighbors only because we find that the magnitude of the interatomic force constants in the LDA dynamical matrix decay very fast beyond the second-nearest neighbor and because such interactions are sufficient to predict the small Davydov splitting in the bulk. The spatial variation of these force constants is shown in Fig. 3 (see figure caption for notation). In general, the out-of-plane compressive force constants are larger in magnitude than the in-plane shear force constants in the $N$TL systems. Thus, the frequencies of out-of-plane modes are higher than those of in-plane modes with the same generic displacements. The inter-TL vdW interaction, captured by $k_{\text{S-S-1}}$, is about one order smaller in magnitude than the second-nearest neighbor S-S interaction within one TL ($k_{\text{S-S-2}}$), two orders smaller than the Mo-S interaction within the TL ($k_{\text{Mo-S-1}}$), and in the same order of the Coulomb interaction between S and Mo in adjacent TLs ($k_{\text{Mo-S-2}}$).

We also see clearly that the force constants between atomic pairs at or closest to the surface are different from those in the interior of thin films. Taking $k^S_{\text{Mo-S-1}}$ in 6 TL as an example, the magnitude of the force constants corresponding to the surface Mo-S bond is ~ 2.6% larger than that in the middle. The larger magnitude of the intralayer force constants at the surface are consistent with the shorter Mo-S bond lengths at the surface compared with the interior of the thin films, as observed in the experiment.[40] From Fig. 3, we also note a clear difference between the compressive force constants from the bulk and the interior of the thin films. For example, the compressive force constant relating Mo and its second-nearest S neighbor $k^C_{\text{Mo-S-2}}$, is about seven times larger in thin films than in bulk. Table 1 indicates that the Born effective charges of Mo and S are quite similar in bulk and 1 TL, suggesting that the difference in force constants may be explained by a decrease of dielectric screening in thin films in the out-of-plane direction, resulting in a stronger long-range Coulomb interaction in few-TL systems. However, since the in-plane periodicity is maintained in thin film systems, the value of shear force constants in the interior of $N$TLs is much closer to that in the bulk.

We further examine the spatial distribution of the self-interaction terms, defined to be the diagonal elements of the dynamical matrix at $\Gamma$ point,[25] as computed within LDA (Fig. 4). Both the in-plane and the out-of-plane components of the self-interaction terms for Mo and S are significantly larger at the surface than in the interior of the thin films. Furthermore, the in-plane self-interaction terms are the same in the interior of the thin films (> 1 TL) as in the bulk, clearly indicating that in $N$-TL ($N > 1$) systems, any difference in these in-plane self-interaction terms with the bulk are attributed completely to surface effects. Since the $E^1_{2g}$ mode with the anomalous frequency trend involves in-plane displacements, this result strongly suggests that surface effects, rather than differences in screening, are responsible for the anomalous trend.

### E. Physical Picture of Frequency Trends in Thin Films

The above indicates that besides the intuitive picture of a larger number of interaction pairs for thicker films, one must also take into account the different force constants related with surface atoms, as well as the different force constants in the thin film compared to those in the bulk, to understand frequency trends. As described in the APPENDIX, we have constructed a force constants model for bulk $MoS_2$ by including interactions up to second-nearest neighbors, and fitting the force constants using LDA frequencies. We now construct a similar force constants model for 1 TL $MoS_2$. There are only two force constants, $k_{Mo-S-1}$ and $k_{S-S-2}$, in 1 TL. The fitted force constants for 1 TL contain the subtle effects of surface contraction in thin film systems. Interestingly, the ratio of $k^S_{Mo-S-1}$ ($k^C_{Mo-S-1}$) in 1 TL to that in the bulk is 1.026 (1.035) in the above-derived force constants model, close to the corresponding ratio of force constants in 1 TL to bulk in our LDA-calculated force constants (Fig. 3), which is 1.024 (1.033). Therefore, to construct force constants models for $N$TL $MoS_2$, we include interactions up to second-nearest neighbors, and specify force constants at the surface and in the interior of the thin films by fixing their ratios to the fitted force constants in the bulk model. These ratios are in turn obtained from the LDA calculations, as displayed in Fig. 3.

To distinguish different physical effects, we compute the frequencies using force constants models that use force constant parameters from (1) the bulk, (2) the interior of the thin films only, and (3) the interior of the thin films and the surface of the thin films (for the surface force constants).

From Fig. 5(b) and 5(e), we see that models 1 and 2 give similar results for the frequency trends. Going from bulk to 1 TL, the $A_{1g}$ frequencies predicted by models 1 and 2 decrease much faster than in the LDA and in the experiment, while the $E_{2g}^1$ frequencies red shift instead of blue shift. Since model 2, with parameters from the interior of the thin films, captures the weaker dielectric screening compared to the bulk (model 1), Figs. 5(b) and 5(e) also indicate that differences in screening have negligible effects on the frequency of the in-plane $E_{2g}^1$ mode while causing a minor red shift of the out-of-plane $A_{1g}$ mode. Including the surface effect in model (3) changes the results significantly. Notably, the $E_{2g}^1$ frequencies predicted by model (3) blue shift from bulk to 1 TL, in excellent quantitative agreement with the LDA frequencies. The corresponding $A_{1g}$ frequencies are also in better quantitative agreement with the experiment. Because most experimental results are obtained using supported samples, we further consider model 4, where only one of the two surfaces has modified surface force constants. The magnitude of the interface force constants for the supported sample might lie between those of the interior and the surface force constants. The trends predicted by model 4 are qualitatively similar to those predicted by model 3 where both surfaces have modified force constants. This finding is consistent with the experimental observation that the trends for suspended samples are the same as those for supported samples.[16,17] The above analysis clearly shows that surface effects are responsible for the anomalous blue shift in the $E_{2g}^1$ frequencies, going from bulk to 2D. Furthermore, the same correct trends can be obtained using force constant models with force constants fitted for nearest-neighbor interactions only, as long as the surface effect is included. The predicted frequencies are essentially unchanged when only the increased surface force constant for $k^S_{Mo-S-1}$ is used (with all other force constants equal to their interior values). This shows that the essential physics responsible for the anomalous blue shift in the $E_{2g}^1$ frequencies arises solely from the effects of stronger Mo-S bonds at the surface, and correspondingly larger in-plane Mo-S force constants. This is in contrast to the physics responsible for the anomalous Davydov splitting in the bulk, where second-nearest neighbours interactions are important.

We can obtain an intuitive understanding of these results by examining the corresponding phonon displacement vectors (Fig. 5(c) and 5(f)). Generally, when the material goes from 3D to 2D with no changes in force constants, the frequency for modes such as the $A_{1g}$ and $E_{2g}^1$ modes, where displacement between adjacent S atoms is out-of-phase, tends to decrease, due to fewer interaction pairs. We call this the thickness effect, which is probed directly in models 1 and 2 above. However, the larger intralayer Mo-S force constants at the surface will increase the frequencies of the thin films, and will have an increasing effect as the film thickness decreases, because of an increasing percentage of surface bonds relative to interior bonds. Why are the frequency trends for the $A_{1g}$ and $E_{2g}^1$ modes different? The interior shear force constants that affect the in-plane $E_{2g}^1$ mode are at least 1/3 smaller than the interior compressive force constants that affect the out-of-plane $A_{1g}$ mode. Therefore, the thickness effect is much larger in the $A_{1g}$ mode than in the $E_{2g}^1$ mode, where the frequency drops by only ~0.6 cm$^{-1}$ from bulk to 1 TL in model 1. The displacement of surface atoms decreases significantly with increasing thickness for the $E_{2g}^1$ mode, but remains similar to the interior displacements in the $A_{1g}$ mode. For instance, the atomic displacement at the edge is about one order smaller than that in the middle in 4 TL for the $E_{2g}^1$ mode, as shown in Fig. 5(f). As a result, increasing surface effects with decreasing thickness dominate over thickness effects for the $E_{2g}^1$ mode, while the thickness effect dominates in the $A_{1g}$ mode.

In addition to the force constants model above, we demonstrate our conclusion that surface effects are responsible for the anomalous frequency trends by performing comparative first-principles calculations. Firstly, we compare the frequencies predicted by different exchange-correlation functionals. It is instructive to note that PBE-D2 predicts only a 0.57 cm$^{-1}$ increase in $E_{2g}^1$ frequency going from bulk to 1 TL, in contrast to LDA that predicts a 4.08 cm$^{-1}$ increase, which is in much better agreement with the experiment (as will be shown in Fig. 7). Analyzing the spatial variation of force constants (Fig. 6), we see that the magnitude of the force constant corresponding to the surface Mo-S bond is only 0.7% larger than the interior Mo-S force constants in PBE-D2, in contrast to a corresponding value of 2.6% in LDA. There is therefore a clear correlation between the relative increase in the surface force constants and the magnitude of the blue shift in $E_{2g}^1$ frequencies. (Our PBE-D2 frequencies, also obtained using PBE-D2 optimized geometries, give the correct trends in both $E_{2g}^1$ and $A_{1g}$ frequencies, in contrast to previous reports on Perdew-Wang 1991 exchange functional with dispersion correction, or PW91+D calculations[24]; Fig. 7)

Secondly, we also compute the "phonon frequencies" for an artificial geometry (called the LDA surface-expanded structure). Specifically, based on the LDA-relaxed structures, we increase the surface Mo-S bond length until the Mo-S force constants at the surface are the same as those in the interior of thin films (Fig. 6). This requires the surface Mo-S bond length to be increased from 2.419Å to 2.426Å in 1- to 5-TL MoS$_2$. Here, we define "phonon frequencies" to be the square root of the eigenvalues of the dynamical matrix, which are well defined even though phonons are not properly defined in the out-of-equilibrium structure. The resulting LDA-computed $E_{2g}^1$ "phonon frequencies" red shift from bulk to 1 TL, indicating clearly that enhanced surface force constants are necessary for the blue shift in $E_{2g}^1$ frequencies. If we use in LDA the bulk Mo-S bond lengths at the surface, the surface Mo-S force constants are still larger than those in the interior (by 1.7% instead of 2.6% for relaxed structures; see Fig. S4 in the supplemental material), and accordingly, the anomalous frequency shifts are still observed, but to a smaller extent than in the fully relaxed structures. Therefore, the "surface effect", i.e. the larger surface force constants, is present even when the bond lengths at the surface are artificially fixed to bulk values. This suggests that the larger surface force constants arise simply because atoms at the surface are short of neighbors and do not require shorter Mo-S surface bonds.

As a previous report[25] had related the anomalous $E_{2g}^1$ frequency trend to the dielectric tensor elements $\varepsilon_{xx}$ and $\varepsilon_{zz}$, it is instructive to compare these quantities in our different first principles calculations. Similar to the procedure in Ref. 25, we compute $\varepsilon_{xx}$ and $\varepsilon_{zz}$ within DFT (Quantum Espresso) for a 16Å vacuum thickness. As expected, $\varepsilon_{xx}$ and $\varepsilon_{zz}$ are essentially the same for LDA, PBE-D2 and LDA-surface-expanded structures (Table SII in the supplemental material), even though the frequency trends for the $E_{2g}^1$ mode are distinctly different. This result clearly indicates that the anomalous frequency trends for the $E_{2g}^1$ mode are due to the surface effect, not due to differences in screening in the bulk and 1 TL. While we know that the surface effect (increased surface force constants) must result from a loss of neighbors in adjacent layers, we also seek to quantify the range of interactions at which this surface effect becomes important. Specifically, we compute the force constants $k^S_{\text{Mo-S-1}}$ and $E_{2g}^1$ frequencies for bulk MoS$_2$ systems while systematically increasing the distance between adjacent MoS$_2$ layers. We find that the magnitude of $k^S_{\text{Mo-S-1}}$ and $E_{2g}^1$ frequencies both increase as the interlayer distance in the bulk is increased by 0 to ~1.5 Å (corresponding to interlayer S-S distances of 2.87 Å to 4.37 Å). Further increase in the interlayer distance does not change the magnitude of $k^S_{\text{Mo-S-1}}$ and $E^1_{2g}$ frequencies significantly, indicating that most of the important interactions resulting in the surface effect are present within 1.5 Å of the equilibrium interlayer distance (see Fig. S5 in the supplemental material).

In addition to MoS$_2$, the analysis given here can be applied to explain the anomalous Raman evolution trends found in the other TMD materials, such as MoSe$_2$,[26,27] WS$_2$,[28] and WSe$_2$.[17,29] Furthermore, the anomalous blue shift in the $E^1_{2g}$ mode also arises directly from the anisotropic bonding in this layered material. For example, the confinement effect in silicon nanocrystals results in a red shift of phonon frequencies, because the lowering of the coordination number at the surface (analogous to the reduction in the number of S-S interlayer pairings in MoS$_2$, i.e. the thickness effect) has a larger effect than the increased surface force constants.[57] For the $E^1_{2g}$ mode in layered MoS$_2$, the thickness effect is negligible, suggesting that 2D layered materials have intrinsically different mechanical properties than typical nanoscale covalent compounds.

## IV. CONCLUSION

Using experiments and first-principles calculations, we have found that the $E^1_{2g}$ mode blue shifts when MoS$_2$ goes from bulk to 2D. We show clearly that this unusual trend arises from the increased surface force constants for Mo-S intra-TL interactions in thin film MoS$_2$. These increased surface force constants arise from the loss of neighbours in adjacent layers. The observed subtle competition between surface effects and thickness effects is likely to be important in other 2D layered materials, such as multi-layer graphene, BN, $Bi_2Te_3$, $Bi_2Se_3$ and $WSe_2$, which all exhibit fascinating and technologically useful properties.

## ACKNOWLEDGEMENTS

We thank L. Wirtz for helpful discussions. S.Y.Q. gratefully acknowledges support from the Institute of High Performance Computing Independent Investigatorship. Q.X. gratefully acknowledges support from the Singapore National Research Foundation through a fellowship grant No. NRF-RF2009-06. This work was also supported in part by Ministry of Education via Tier 2 grant No. MOE2012-T2-2-086 and start-up grant support No. M58113004 from Nanyang Technological University. The authors acknowledge the support from the A*STAR Computational Resource Center.

## APPENDIX

### 1. Force constants model for bulk $MoS_2$

The equation of motion for the normal vibrations of atom $n$ in a linear chain representation can be expressed as:

$$M_n \frac{d^2 u_n}{dt^2} = -\sum_m k_{nm}(u_m - u_n) \tag{A1}$$

where $M_n$ is the mass of atom $n$; $k_{nm}$ is the force constant between atoms $n$ and $m$; $u_n$ and $u_m$ are the normal displacement of atoms $n$ and $m$, respectively; and the displacement is assumed to have the form $u_n(x,t) = v_n(x)\exp(-i\omega t)$. Refer to Fig. 1b of main text for the numbering of the atoms in the bulk $MoS_2$ unit cell, and the labeling of force constants. The phonon frequencies can be obtained by diagonalizing the corresponding dynamical matrix. In the present force constants model, only the interactions between the first- and second-nearest neighbors are considered; therefore, we have four force constants in the final equations. In this paper, $k_{Mo\text{-}S\text{-}1}$ represents the force constants between Mo and its nearest S atoms, and $k_{S\text{-}S\text{-}1}$ represents the interaction between neighboring S atoms in adjacent TLs. $k_{S\text{-}S\text{-}1}$ is very small compared with the $k_{Mo\text{-}S\text{-}1}$ due to the weak vdW interaction between adjacent TLs. For the second-nearest neighbors' interaction, $k_{Mo\text{-}S\text{-}2}$ is the force constant between Mo and its second-nearest S atoms, and $k_{S\text{-}S\text{-}2}$ describes the interaction between S and S in the same TL. The force constants can be further divided into the shear and compressive force constants, depending on the direction of atomic displacements (in plane or out of plane).

According to group theory, bulk MoS$_2$ belongs to space group $D_{6h}^4$ ($P6_3/mmc$).[48] The primitive unit cell has six atoms, resulting in 18 Brillouin Zone center (Γ) phonons with 3 acoustic phonon modes and 15 optical phonon modes. The irreducible representations of the phonon modes at $\Gamma$ point can be decomposed into $\Gamma = A_{1g} + 2A_{2u} + B_{1u} + 2B_{2g} + E_{1g} + 2E_{1u} + E_{2u} + 2E_{2g}$. The in-plane modes are twofold degenerate. From the symmetry of atomic eigenvectors displacements, the 18 phonon modes can be divided into four groups:[50]

1. 2 $A_{2u}$ and 2 $E_{1u}$: the displacements of eigenvectors follow $u_1 = u_3 = u_4 = u_6$, and $u_2 = u_5$.

2. 2 $B_{2g}$ and 2 $E_{2g}$: the eigenvectors for these modes have the relations: $u_1 = u_3 = -u_4 = -u_6$, and $u_2 = -u_5$.

3. $A_{1g}$ and $E_{1g}$: the Mo atoms do not move ($u_2 = u_5 = 0$), and $u_1 = -u_3 = u_4 = -u_6$.

4. $B_{1u}$ and $E_{2u}$: the Mo atoms do not move ($u_2 = u_5 = 0$), and $u_1 = -u_3 = -u_4 = u_6$.

The first- and second-nearest neighbor interactions are considered by substituting the four force constants $k_{Mo\text{-}S\text{-}1}$, $k_{S\text{-}S\text{-}1}$, $k_{Mo\text{-}S\text{-}2}$, and $k_{S\text{-}S\text{-}2}$ into Eq. (A1). By using the symmetry of the atomic displacements in each phonon mode mentioned above, we can get four groups of corresponding frequencies in the expression of the force constants:

1) $\omega^2(A_{2u}^1, E_{1u}^1) = 0$

$$\omega^2(A_{2u}^2, E_{1u}^2)G = \frac{(2m_s + m_{mo})(k_{Mo-S-1} + k_{Mo-S-2})}{m_s m_{mo}} \qquad (A2)$$

where, $G = -2\pi^2 c^2 \mu$ is a factor, $c$ is the speed of light in cm s$^{-1}$, $\mu$ is the atomic mass per unit area, $m_s$ and $m_{mo}$ are the masses of Mo and S atoms, $k_{Mo-S-1} + k_{Mo-S-2}$ can be considered as the effective force constant of Mo and S.

2) 
$$\omega^2(B_{2g}^1, E_{2g}^1)G = \frac{2k_{S-S-1} + (k_{Mo-S-1} + k_{Mo-S-2})}{2m_s} + \frac{(k_{Mo-S-1} + k_{Mo-S-2})}{m_{mo}} +$$
$$\sqrt{\left(\frac{2k_{S-S-1} + (k_{Mo-S-1} + k_{Mo-S-2})}{2m_s} - \frac{(k_{Mo-S-1} + k_{Mo-S-2})}{m_{mo}}\right)^2 + \frac{2(k_{Mo-S-1} - k_{Mo-S-2})^2}{m_s m_{mo}}}$$

$$\omega^2(B_{2g}^2, E_{2g}^2)G = \frac{2k_{S-S-1} + (k_{Mo-S-1} + k_{Mo-S-2})}{2m_s} + \frac{(k_{Mo-S-1} + k_{Mo-S-2})}{m_{mo}} -$$
$$\sqrt{\left(\frac{2k_{S-S-1} + (k_{Mo-S-1} + k_{Mo-S-2})}{2m_s} - \frac{(k_{Mo-S-1} + k_{Mo-S-2})}{m_{mo}}\right)^2 + \frac{2(k_{Mo-S-1} - k_{Mo-S-2})^2}{m_s m_{mo}}}$$

(A3)

From $\dfrac{\omega^2(B_{2g}^1, E_{2g}^1) + \omega^2(B_{2g}^2, E_{2g}^2)}{2} G = \dfrac{2k_{S-S-1} + (k_{Mo-S-1} + k_{Mo-S-2})}{2m_s} + \dfrac{(k_{Mo-S-1} + k_{Mo-S-2})}{m_{mo}}$, $k_{S-S-1}$

can be readily obtained when combined with Eq. (A2). With the obtained $k_{S-S-1}$, $k_{Mo-S-1}$ and $k_{Mo-S-2}$ can also be calculated separately with the help of Eqs. (A3) and (A2).

3) $\omega^2(A_{1g}, E_{1g})G = \dfrac{(k_{Mo-S-1} + k_{Mo-S-2}) + 2k_{S-S-2} + 2k_{S-S-1}}{m_s}$ (A4)

4) $\omega^2(B_{1u}, E_{2u})G = \dfrac{(k_{Mo-S-1} + k_{Mo-S-2}) + 2k_{S-S-2}}{m_s}$ (A5)

From Eqs. (A2) and (A5), we can get the force constant $k_{S-S-2}$.

By substituting the LDA-calculated frequencies, we can get four shear force constants and four compressive force constants in our models, as shown in Table AI.

## 2. Force constants model for 1 TL MoS$_2$

In the 1 TL MoS$_2$ thin film system, we can still apply Eq. (A1). Since there are only three atoms in the 1 TL, only the force constants $k_{Mo-S-1}$ and $k_{S-S-2}$ exist. With the help of symmetry of the phonon modes in the space group $D_{3h}^1$ ($P\bar{6}m2$) for 1 TL,[17] we obtain, following the same procedure as before, the following:

1) $A_2'' + E'$: the displacements have the relationship $u_1 = u_3$ and $u_2 \neq 0$.

$\omega^2(A_2'', E')G = \dfrac{(2m_s + m_{mo})k_{Mo-S-1}}{m_s m_{mo}}$ (A6)

2) $A_1' + E''$: $u_1 = u_3$ and $u_2 = 0$.

$$\omega^2(A_1', E'')G = \frac{k_{Mo-S-1} + 2k_{S-S-2}}{m_s} \tag{A7}$$

Thus, we can get the following force constants for the 1 TL MoS$_2$. All the parameters are summarized in Table AI.

Table AI. Parameters derived from fitting the force constants model to the LDA frequencies. The unit is N/m$^3$.

|      |              | $k_{Mo-S-1}$ | $k_{Mo-S-2}$ | $k_{S-S-1}$ | $k_{S-S-2}$ |
|------|--------------|--------------|--------------|-------------|-------------|
| Bulk | In-plane     | -9.614×10$^{20}$ | -6.86×10$^{18}$ | -3.57×10$^{18}$ | 4.27×10$^{19}$ |
|      | Out of plane | -14.039×10$^{20}$ | -4.43×10$^{18}$ | -37.4×10$^{18}$ | -18.21×10$^{19}$ |
| 1 TL | In-plane     | -9.848×10$^{20}$ | - | - | 4.40×10$^{19}$ |
|      | Out of plane | -14.531×10$^{20}$ | - | - | -16.87×10$^{19}$ |

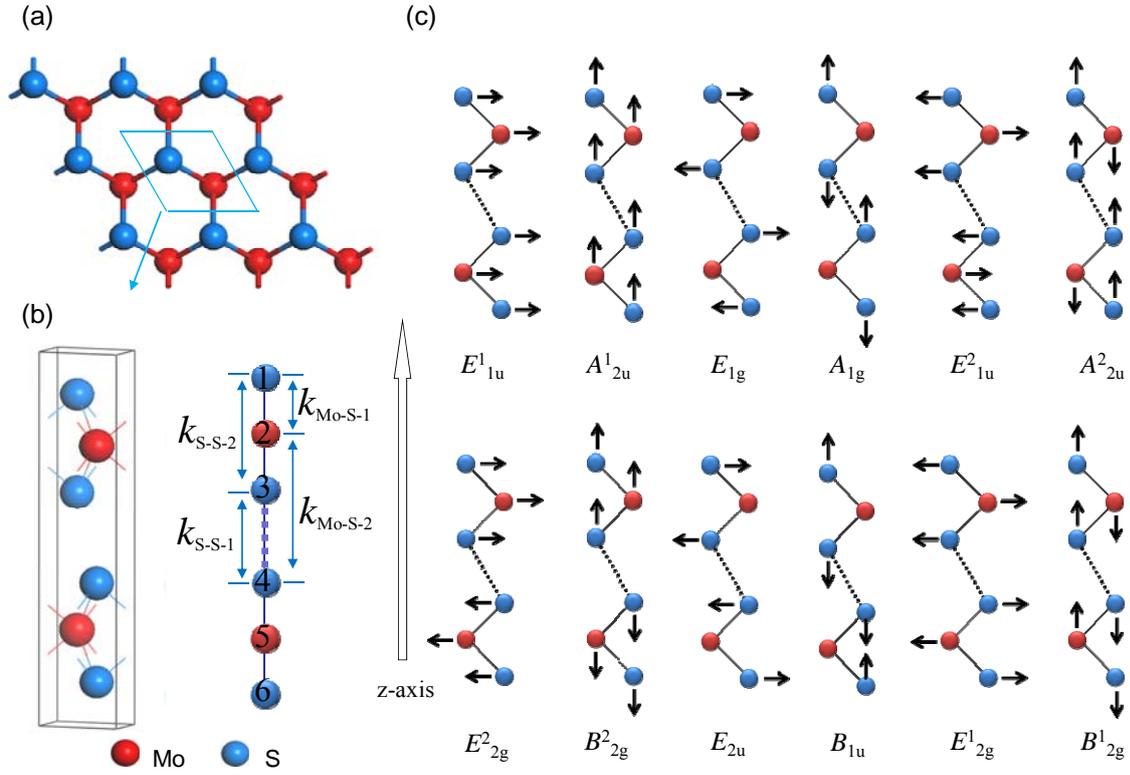

**Figure 1.** (Color online) (a) Top view of atoms of bulk 2H-MoS$_2$; the primitive unit cell runs over two TLs and is shown by the blue diamond. (b) Side view of the unit cell and its linear chain representation with force constants up to the second-nearest neighbors. The notation for the force constants is explained in the caption for Fig. 3. (c) Displacement representations for the $\Gamma$-point phonon vibrational modes. Each mode in the top row forms a conjugate pair with the mode vertically below it in the bottom row.

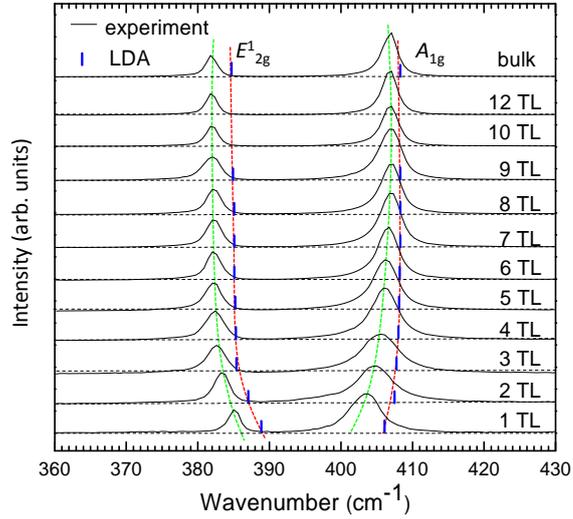

**Figure 2.** (Color online) **Experimental and LDA-calculated frequencies for the $E^1_{2g}$ and $A_{1g}$ modes in few-TL and bulk MoS$_2$.** The $E^1_{2g}$ modes blue shift and the $A_{1g}$ modes red shift as the thickness of the MoS$_2$ film decreases. The green and red dashed lines are used as a guides to the trends for the experimental and LDA results, respectively.

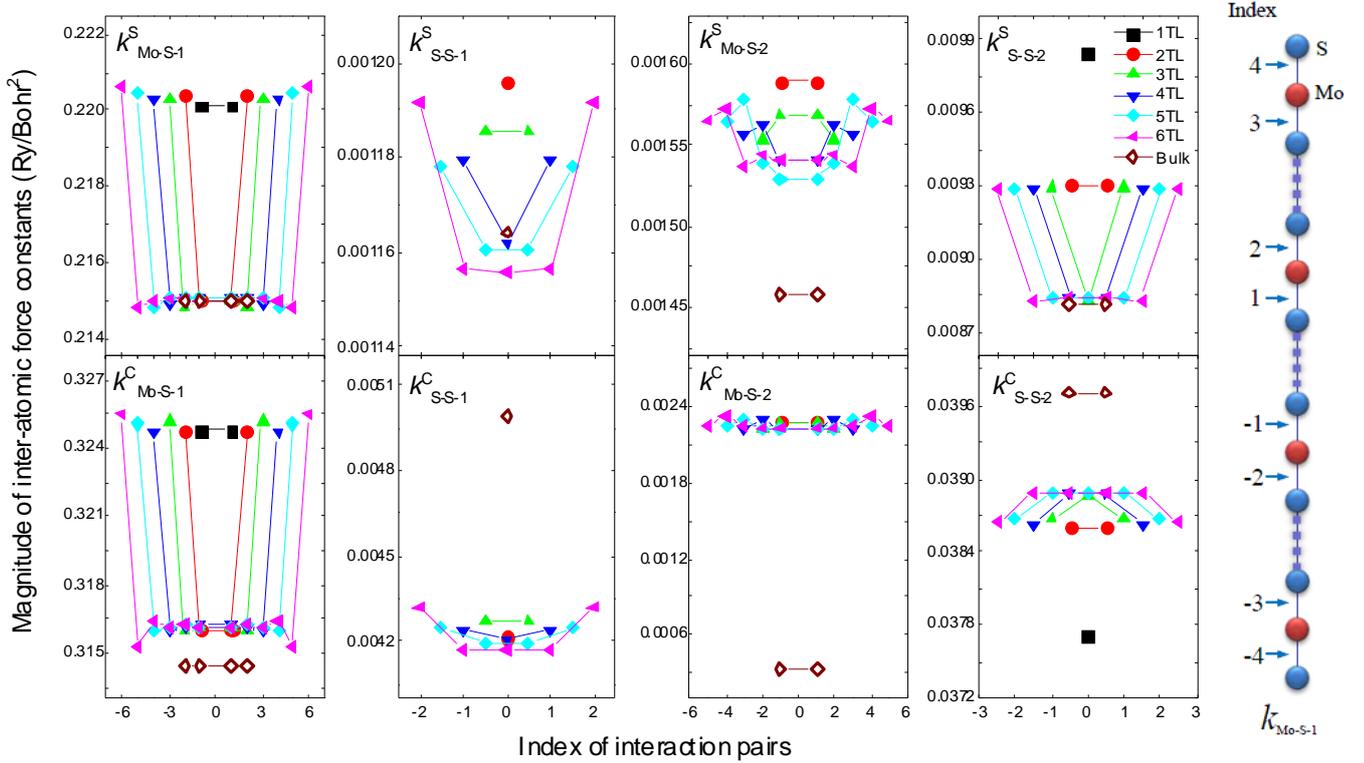

**Figure 3.** (Color online) **Spatial variations of the magnitude of LDA interatomic force constants in bulk and $N$TL MoS$_2$.** The legend is in the top right panel. The superscript, 'S' or 'C' in the force constant, refers to the in-plane 'shear' or out-of-plane 'compressive' force constant, respectively. The subscript refers to the elements of the atoms in the interatomic pair for which the force constant is computed, followed by a number indicating whether these atoms are first-nearest ("1") or second-nearest ("2") neighbors. The index of interaction pairs (horizontal axis) is counted from the bottom TL to the surface TL, with the middle set to 0. An illustration of the index of $k_{\text{Mo-S-1}}$ in 4 TL MoS$_2$ in the linear chain representation is shown on the right-hand side. For example, an index of '-4' refers to the force constant between the bottommost S and Mo atoms in 4 TL MoS$_2$.

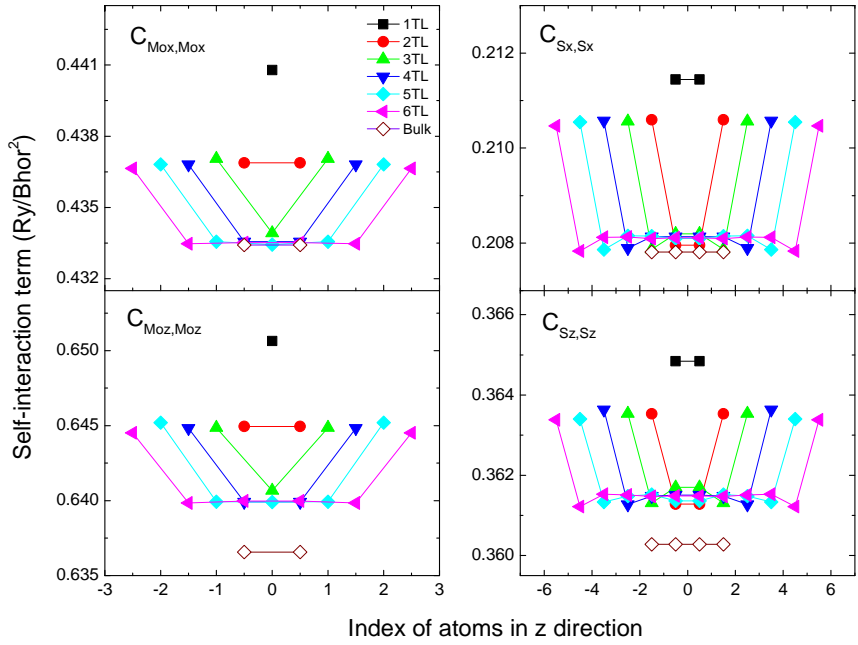

**Figure 4. (Color online) Spatial variations of the LDA-calculated self-interaction terms of Mo and S atoms in fully optimized 1-6 TL and bulk MoS$_2$.** $C_{Mox,Mox}$ and $C_{Moz,Moz}$ represents the in-plane *xx* component and out-of-plane *zz* component of the self-interaction term in Mo atoms. A similar definition applies to $C_{Sx,Sx}$ and $C_{Sz,Sz}$ for S atoms.

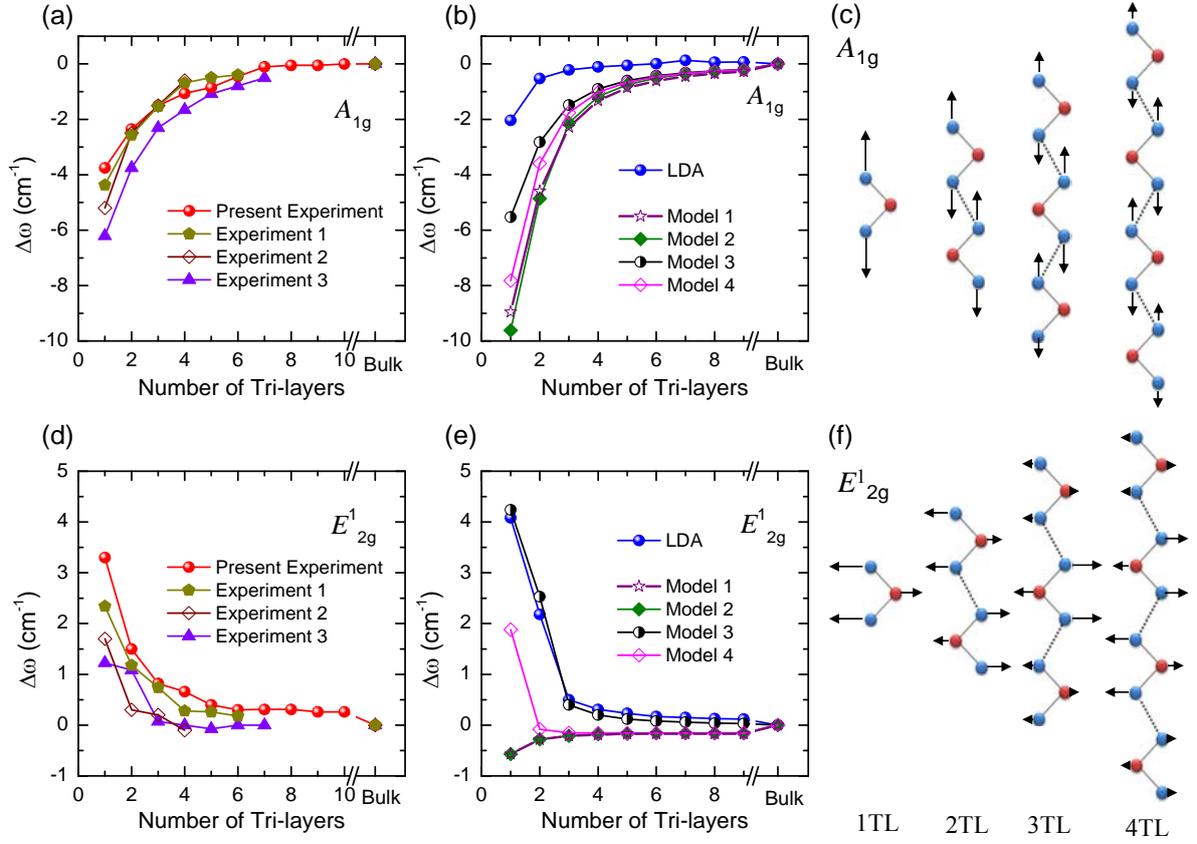

**Figure 5. (Color online) Frequency evolutions of the high-frequency $A_{1g}$ (a,b,c) and $E^1_{2g}$ (d,e,f) counterpart modes in *NTL* and bulk MoS$_2$.** Only the frequency differences with respect to their corresponding bulk values are displayed. Data of experiments 1, 2 and 3 are from Ref 16, 22, and 20 respectively. The force constants model results for $A_{1g}$ and $E^1_{2g}$ modes are shown in (b) and (e). As described in the text, model 1 uses parameters from the bulk, model 2 uses parameters from the interior of the thin films, model 3 is the same as model 2 but taks into account modified surface force constants on both surfaces, and model 4 is like model 3 but with surface force constants on one surface only.

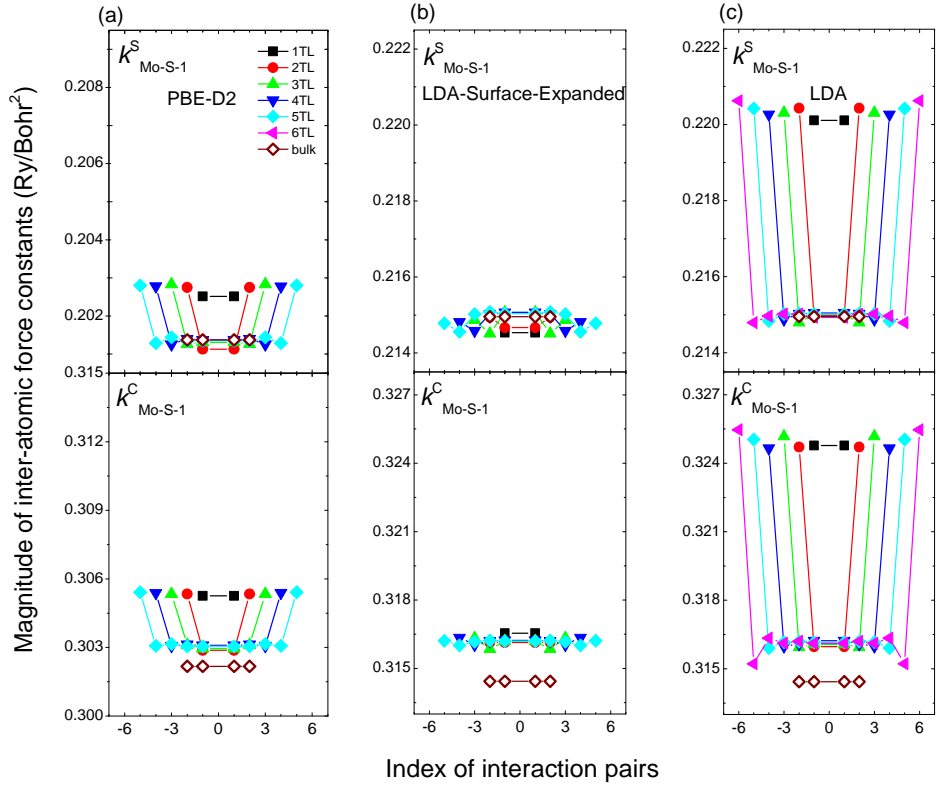

**Figure 6. (Color online) Spatial variations of the magnitude of $k_{\text{Mo-S-1}}$ force constants in bulk and 1-5 TL for (a) PBE-D2 optimized structures and (b) LDA-surface-expanded systems**. For the purpose of comparison, the corresponding values of LDA optimized structures are also provided in (c). The same span is used in the y-axis. The force constants corresponding to other atomic pairs are very similar for all three calculations, as we show in Figs. S2 and Fig. S3 (see supplemental material).

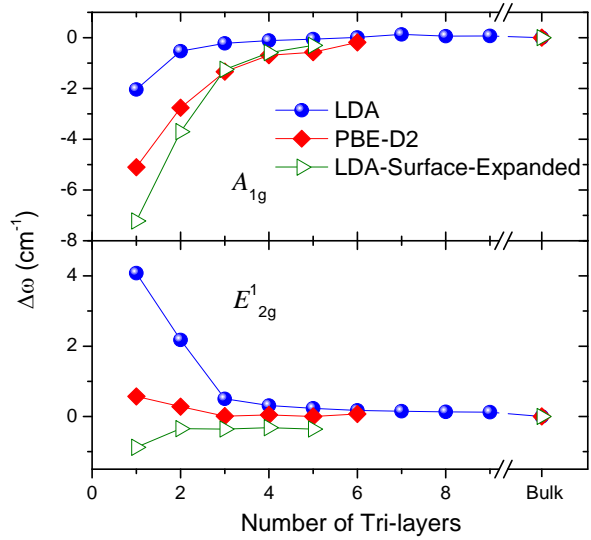

**Figure 7. (Color online) Frequency evolution of the $A_{1g}$ and $E^1_{2g}$ modes in $N$TL and bulk MoS$_2$ for LDA and PBE-D2 optimized structures, the LDA calculated "phonon eigenvalues" of surface-expanded structures are shown in open triangles.** Only the frequency differences with respect to their corresponding bulk values are displayed.

Table I. Lattice constants ($a$, $c$), bond length ($d_{\text{Mo-S}}$, $d_{\text{S-S}}$) within each TL, and Born effective charge $Z^*$ of bulk and 1 TL MoS$_2$.

|  |  | $a$ (Å) | $c$ (Å) | $d_{\text{Mo-S}}$ | $d_{\text{S-S}}$ | $Z^*(Mo)$ | $Z^*(S)$ |
|---|---|---|---|---|---|---|---|
| LDA | bulk | 3.169 | 12.086 | 2.421 | 3.171 | 1.37 | -0.68 |
|  | 1 TL | 3.167 | - | 2.419 | 3.168 | 1.39 | -0.69 |
| PBE-D2 | bulk | 3.198 | 12.426 | 2.418 | 3.121 | 0.95 | -0.50 |
|  | 1 TL | 3.200 | - | 2.419 | 3.123 | 0.96 | -0.45 |
| vdW-DF | bulk | 3.191 | 12.239 | 2.439 | 3.197 | 1.23 | -0.53 |
|  | 1 TL | 3.186 | - | 2.435 | 3.193 | 1.26 | -0.56 |

**Table II.** Calculated $\Gamma$ point phonon frequencies (in cm$^{-1}$) of 2H bulk and 1 TL MoS$_2$ with different functionals and using the force constants model including interactions up to the second-nearest neighbors, compared to the experiment.

| MoS$_2$ | Irr. rep. | Exp. | LDA | PBE-D2 | vdW-DF | model |
|---|---|---|---|---|---|---|
| Bulk | $E^1_{1u}$ (I) | | 0 | 0 | 0 | 0 |
| | $A^1_{2u}$ (I) | | 0 | 0 | 0 | 0 |
| | $E^2_{2g}$ (R) | 33.7$^a$, 31.8$^e$ | 35.69 | 27.08 | 36.40 | 35.69 |
| | $B^2_{2g}$ | 56$^b$ | 57.94 | 51.81 | 68.80 | 57.94 |
| | $E_{2u}$ | | 285.16 | 276.11 | 279.98 | 285.16 |
| | $E_{1g}$ (R) | 287$^c$ | 287.78 | 277.83 | 283.00 | 286.31 |
| | $E^1_{2g}$ (R) | 383$^c$, 382.0$^d$, 381.8$^e$ | 384.92 | 372.56 | 378.07 | 384.92 |
| | $E^2_{1u}$ (I) | 384$^c$ | 385.72 | 372.74 | 378.33 | 385.72 |
| | $B_{1u}$ | | 404.03 | 396.77 | 397.76 | 404.03 |
| | $A_{1g}$ (R) | 409$^c$, 407.4$^d$, 407.1$^e$ | 408.11 | 401.71 | 406.56 | 412.47 |
| | $A^2_{2u}$ (I) | 470$^c$ | 465.18 | 455.95 | 458.36 | 465.18 |
| | $B^1_{2g}$ | | 468.96 | 459.46 | 463.88 | 468.96 |
| 1 TL | $E'$ (I+R) | | 0 | 0 | 0 | 0 |
| | $A''_2$ (I) | | 0 | 0 | 0 | 0 |
| | $E''$ (R) | | 287.38 | 276.07 | 287.61 | 287.38 |
| | $E'$ (I+R) | 384.3$^d$, 385.1$^e$ | 389.00 | 373.13 | 389.36 | 389.00 |
| | $A'_1$ (R) | 403.0$^d$, 403.3$^e$ | 406.07 | 396.61 | 404.47 | 406.07 |
| | $A''_2$ (I) | | 472.52 | 458.1 | 472.90 | 472.52 |

$^a$Ref.42

$^b$Ref.41

$^c$Ref.43

$^d$Ref.16

$^e$Our experimental data.

**Table III. LDA calculated $\Gamma$ point phonon frequencies (cm$^{-1}$) and relative Raman intensities $I_{xx}$ (in parentheses) of the phonon modes in MoS$_2$, with intensity is normalized by the biggest value in each row. The irreducible representation and Raman [R] /Infrared [I] activity are also indicated; I+R indicates the mode is both Raman and Infrared active while Ina indicates that the mode is optically inactive.** The background colors show three different branches of vibration modes originated from 1 TL. Numbers in blue, pink and red refer to frequencies that are observed in experiment.

| | | | | | | | | | | | | |
|---|---|---|---|---|---|---|---|---|---|---|---|---|
| **In-plane Modes** | bulk | 0 (0) $E^1_{1u}$ [I] | 35.7 (0.31) $E^2_{2g}$ [R] | 285.2 (0) $E_{2u}$ [Ina] | 287.8 (0) $E_{1g}$ [R] | 384.9 (1.0) $E^1_{2g}$ [R] | 385.7 (0) $E^2_{1u}$ [I] | | | | | |
| | 1 TL | 0 (0) $E'$ [I+R] | 287.4 (0) $E''$ [R] | 389.0 (1.0) $E'$ [I+R] | | | | | | | | |
| | 2 TL | 0 (0) $E_u$ [I] | 25.8 (0.06) $E_g$ [R] | 286.4 (0) $E_u$ [I] | 287.8 (0.02) $E_g$ [R] | 387.14 (1.0) $E_g$ [R] | 387.5 (0) $E_u$ [I] | | | | | |
| | 3 TL | 0 (0) $E'$ [I+R] | 18.3 (0) $E''$ [R] | 31.3 (0.24) $E'$ [I+R] | 285.9 (0) $E''$ [R] | 287.0 (0.02) $E'$ [I+R] | 287.8 (0) $E''$ [R] | 385.4 (1.0) $E'$ [I+R] | 387.19 (0) $E''$ [R] | 387.23 (0.76) $E'$ [I+R] | | | |
| | 4 TL | 0 (0) $E_u$ [I] | 13.7 (0.03) $E_g$ [R] | 25.4 (0) $E_u$ [I] | 33.3 (0.29) $E_g$ [R] | 285.7 (0) $E_u$ [I] | 286.5 (0.006) $E_g$ [R] | 287.4 (0) $E_u$ [I] | 287.9 (0.006) $E_g$ [R] | 385.2 (1.0) $E_g$ [R] | 385.6 (0) $E_u$ [I] | 387.27 (0.42) $E_g$ [R] | 387.28 (0) $E_u$ [I] |
| **Out-of-plane Modes** | bulk | 0 (0) $A^1_{2u}$ [I] | 57.9 (0.0) $B^2_{2g}$ [Ina] | 404.0 (0) $B_{1u}$ [Ina] | 408.1 (1.0) $A_{1g}$ [R] | 465.2 (0) $A^2_{2u}$ [I] | 469.0 (0) $B^1_{2g}$ [Ina] | | | | | |
| | 1 TL | 0 (0) $A''_2$ [I] | 406.1 (1.0) $A'_1$ [R] | 472.5 (0) $A''_2$ [I] | | | | | | | | |
| | 2 TL | 0 (0) $A_{2u}$ [I] | 41.4 (0.46) $A_{1g}$ [R] | 405.5 (0) $A_{2u}$ [I] | 407.6 (1.0) $A_{1g}$ [R] | 470.2 (0) $A_{2u}$ [I] | 471.0 (0.001) $A_{1g}$ [R] | | | | | |
| | 3 TL | 0 (0) $A''_2$ [I] | 28.8 (0.55) $A'_1$ [R] | 51.5 (0) $A''_2$ [I] | 404.9 (0.003) $A'_1$ [R] | 406.5 (0) $A''_2$ [I] | 407.9 (1.0) $A'_1$ [R] | 468.3 (0) $A''_2$ [I] | 470.62 (0.001) $A'_1$ [R] | 470.61 (0) $A''_2$ [I] | | | |
| | 4 TL | 0 (0) $A_{2u}$ [I] | 22.3 (0.53) $A_{1g}$ [R] | 41.5 (0) $A_{2u}$ [I] | 54.0 (0.02) $A_{1g}$ [R] | 404.6 (0) $A_{2u}$ [I] | 405.8 (0.01) $A_{1g}$ [R] | 407.1 (0) $A_{2u}$ [I] | 408.0 (1.0) $A_{1g}$ [R] | 467.9 (0) $A_{2u}$ [I] | 468.8 (0.0001) $A_{1g}$ [R] | 470.6 (0) $A_{2u}$ [I] | 470.8 (0.0003) $A_{1g}$ [R] |

## Anomalous Frequency Trends in MoS$_2$ Thin Films Attributed to Surface Effects

Xin Luo[1], Yanyuan Zhao[2], Jun Zhang[2], Qihua Xiong[2,3,*], Su Ying Quek[1,†,*]

[1]Institute of High Performance Computing, 1 Fusionopolis Way, #16-16 Connexis, Singapore 138632

[2]Division of Physics and Applied Physics, School of Physical and Mathematical Sciences, Nanyang Technological University, 21 Nanyang Link, Singapore 637371

[3]Division of Microelectronics, School of Electrical and Electronic Engineering, Nanyang Technological University, Singapore, 639798

*: To whom correspondence should be addressed. Email address: qihua@ntu.edu.sg (experiment), phyqsy@nus.edu.sg (theory)
†: Present address: Department of Physics, National University of Singapore, S12-M01, 2 Science Drive 3, Singapore 117551

**Contents**

**- Convergence of dielectric tensors in thin films**

**- Comparison of dielectric tensors for different systems**

**- Discussion of phonon line-widths observed in experiment (+ Figure S1)**

**- Figures S2, S3 and S4 showing spatial variations of the magnitude of force constants in, respectively, PBE-D2 systems, "LDA-surface-expanded" systems, and systems in which the LDA surface Mo-S bond lengths are fixed to those in the bulk**

**- Figure S5 shows the frequency of E$^1_{2g}$ mode and the magnitude of $k^S_{Mo\text{-}S\text{-}1}$ as a function of the interlayer S-S distance.**

**Convergence of dielectric tensors in thin films**

In the calculation of dielectric tensors in thin film MoS$_2$, we note that the dielectric tensor is sensitive to the vacuum thickness, as shown in Table SI. We do not address this effect as it is beyond the scope of our present work. Furthermore, we note that our argument that the surface effect is responsible for the anomalous frequency trend does not rely on a quantitatively accurate calculation of the dielectric tensor.

Table SI. The influence of the vacuum thickness (in Å) on the dielectric tensors, $\varepsilon_{xx}$ and $\varepsilon_{zz}$.

| MoS2 | $\varepsilon_{xx}$ | $\varepsilon_{zz}$ |
|---|---|---|
| 1TL(20Å) | 4.507 | 1.286 |
| 1TL(16Å) | 5.243 | 1.368 |
| 1TL(10Å) | 7.137 | 1.637 |
| 1TL (7 Å) | 8.605 | 1.937 |
| bulk | 14.965 | 7.763 |

**Dielectric tensors in three ab initio calculated systems**

Table SII. Comparison of the dielectric tensor elements $\varepsilon_{xx}$ and $\varepsilon_{zz}$ for bulk and 1-2 TL MoS$_2$ (calculated with a given vacuum thicknesses of 16 Å) with different structures.

| MoS2 | LDA | | | PBE-D2 | | | LDA-Surface-Expanded | | |
|---|---|---|---|---|---|---|---|---|---|
| | Bulk | 1TL | 2TL | Bulk | 1TL | 2TL | Bulk | 1TL | 2TL |
| $\varepsilon_{xx}$ | 14.97 | 5.24 | 7.57 | 15.33 | 5.58 | 8.24 | 14.97 | 5.26 | 7.59 |
| $\varepsilon_{zz}$ | 7.76 | 1.37 | 1.70 | 6.12 | 1.36 | 1.73 | 7.76 | 1.37 | 1.71 |

**Discussion of phonon line-widths observed in experiment**

As seen from the Figure 2 of our manuscript, the $A_{1g}$ mode shows a broadening from the bulk to 2D, while a distinct broadening cannot be seen for the $E_{2g}^1$ mode. To make the discussion quantitative, we did double-Lorentzian peak fitting for the $E_{2g}^1$ and $A_{1g}$ peaks for MoS$_2$ layers of different thicknesses and extract the full width half maximum (FWHM) of the peaks. Figure S1 plots out the FWHMs of both $E_{2g}^1$ and $A_{1g}$ peaks as a function of MoS$_2$ layer thickness. The line width of the $A_{1g}$ mode monotonically increases from bulk (~2.6 cm$^{-1}$) to 2TL (~6.4 cm$^{-1}$), and the $E_{2g}^1$ peak also shows a slight broadening from bulk to 3TL, although not as much. The broadening trends with decreasing thickness, however, stop at around 2TL (considering the experimental and fitting errors) with opposite evolution trends from 2TL to 1TL. These results are very similar with what Lee *et al.* reported before (ACS Nano 4, 5 2695 (2010)).

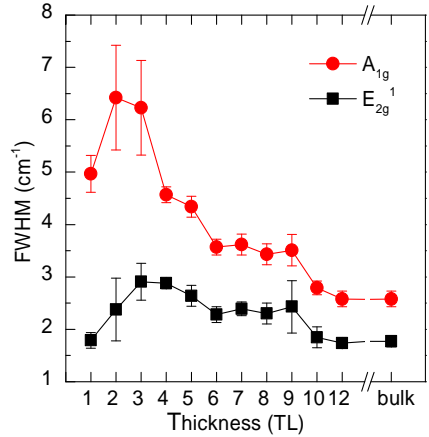

**Figure S1. FWHMs of the $E_{2g}^1$ and $A_{1g}$ modes as a function of MoS$_2$ layer thickness.**

The overall broadening effect of the $E_{2g}^1$ and $A_{1g}$ peaks from bulk to 2D can be well explained by the phonon confinement effect and the uncertainty rule (See Solid State Communications, 39, 625 (1981)). The uncertainty rule relates the position and momentum, $\Delta x \cdot \Delta p \approx h$. Here $\Delta p \approx (h/2\pi)\Delta q$, $\Delta x$ is the material size. In the bulk limit, $\Delta x \to \infty$ and $\Delta q \to 0$, so only zone-center (Γ point) phonons can be experimentally observed. When both $\Delta x$ and $\Delta q$ have finite values, phonons in the vicinity of Γ point will also get involved in Raman scattering, causing a phonon frequency shift and line width broadening, and this is known as the phonon confinement effect. The broadening of a Raman peak also corresponds to a decreased phonon lifetime in 2D, in which case the scattering process involves phonons with different phases and momentums, instead of only Γ phonons in the bulk. The phonon confinement effect gets more dominant when the sample size gets closer to the 2D limit. The large thickness dependence of the $A_{1g}$ line width is related to its out-of-plane vibration manner, along which direction the size of the sample is largely confined. Meanwhile, it is not difficult to understand why the $E_{2g}^1$ mode is less affected considering the fact that the atoms are vibrating in plane and the lateral size of the sample is not necessarily confined from bulk to 2D.

From 2TL to the real 2D limit of 1TL, the Raman peaks are narrowed instead of broadened, not following the phonon confinement effect. We attribute this anomalous phenomenon to the special electronic structure of 1TL $MoS_2$. As has been intensively studied and demonstrated, the band gap of $MoS_2$ (and a few other related transition metal dichalcogenides) experiences an indirect to direct transition at the case of monolayer. Thus, the phonon-electron interactions in 1TL will be very different from 2TL and above. For example, in the indirect band gap cases (2TL and above), phonons with large momentum are largely involved in the photon absorption, emission and phonon scattering processes. While in 1TL, large momentum phonons would not be involved as much given the fact of a direct band gap system.

As discussed above, the scattering of phonons with different momentums (and thus phases) would result in a reduced lifetime, thus it is reasonable to assume that the lifetime of phonons in 1TL should be longer than 2TL and above. The line width evolution of $E_{2g}^1$ and $A_{1g}$ modes is determined by both the phonon confinement effect and the involvement of large momentum phonons during the scattering process. The latter should be less dominant in 1TL, thus giving smaller line widths in 1TL than in 2TL.

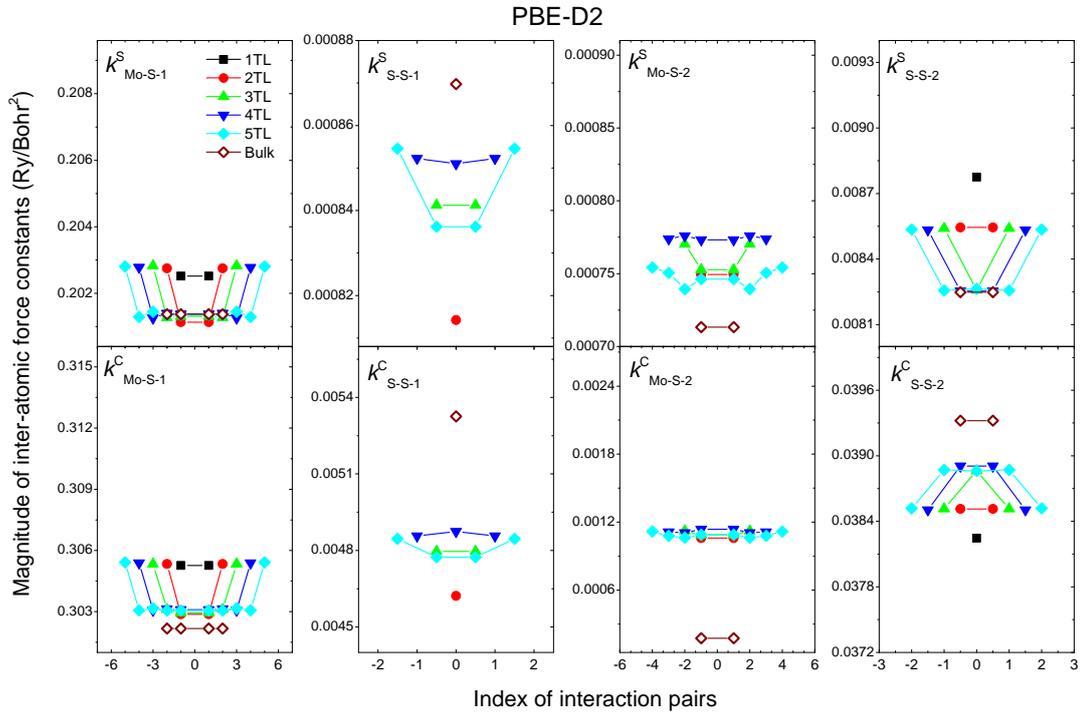

**Figure S2. Spatial variations of the magnitude of inter-atomic force constants for *1-5*TL and bulk MoS$_2$ calculated with PBE-D2.** The legend and notations are the same as Figure 3 of the main text.

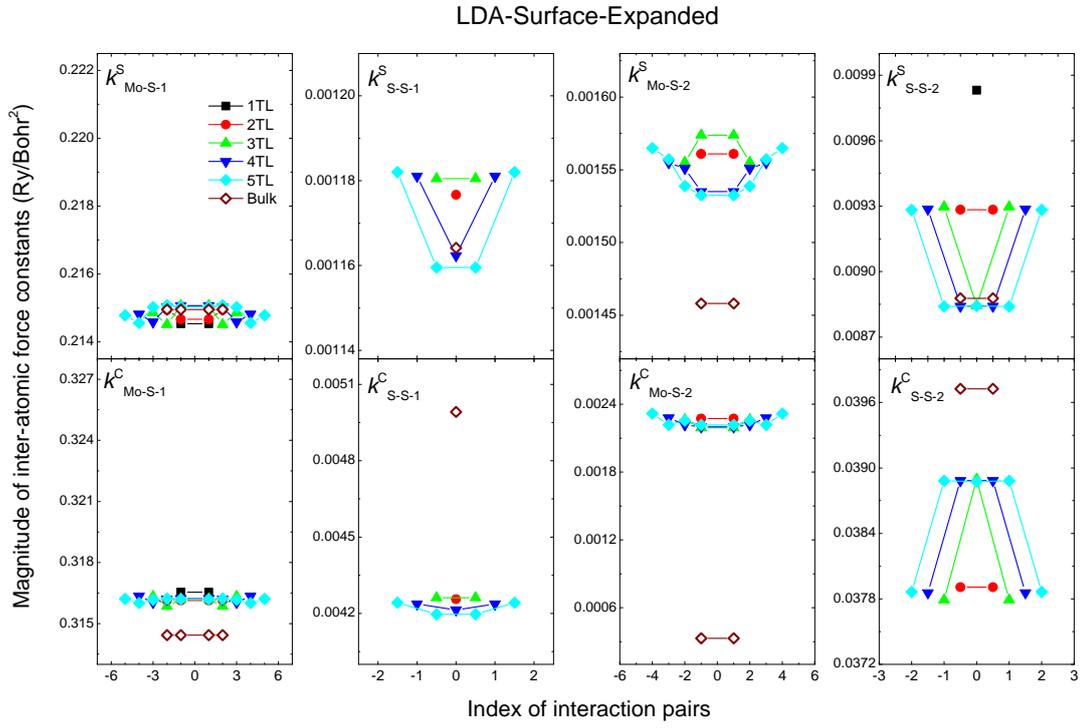

**Figure S3. Spatial variations of the magnitude of LDA calculated inter-atomic force constants in optimized bulk and surface expanded 1-5 TL MoS$_2$.** The legend and notations are the same as Figure 3 of the main text.

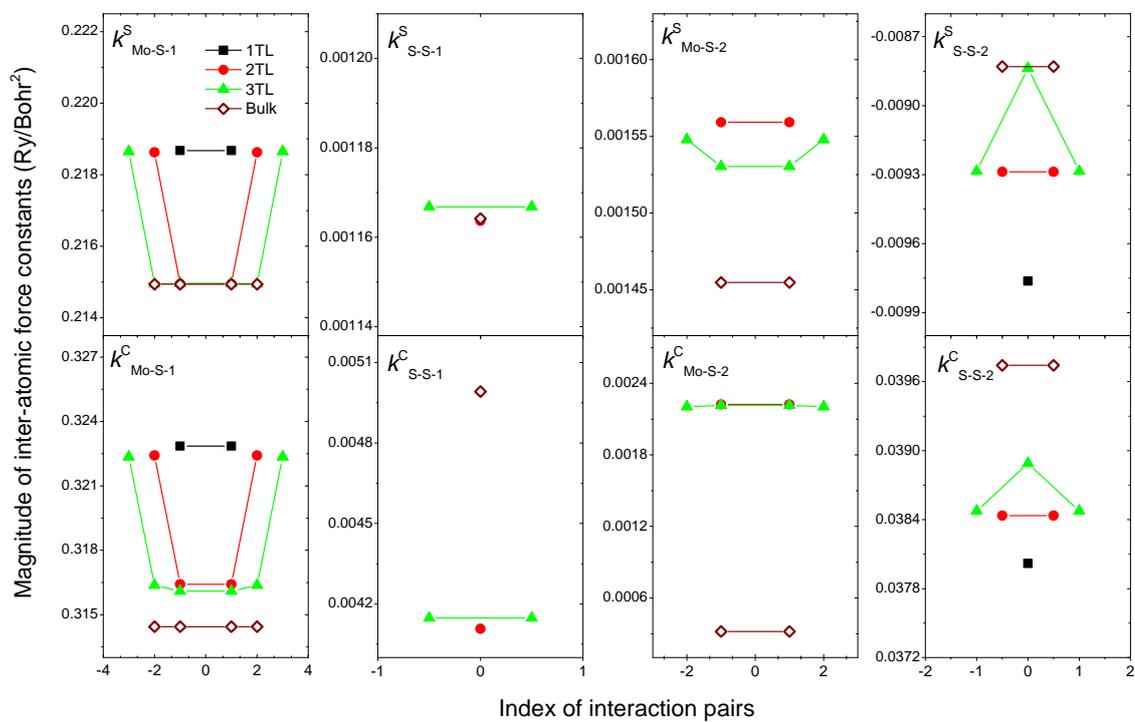

**Figure S4.** Spatial variations of the magnitude of the LDA inter-atomic force constants for systems in which the inter-atomic bond lengths are fixed to those optimized in the bulk.

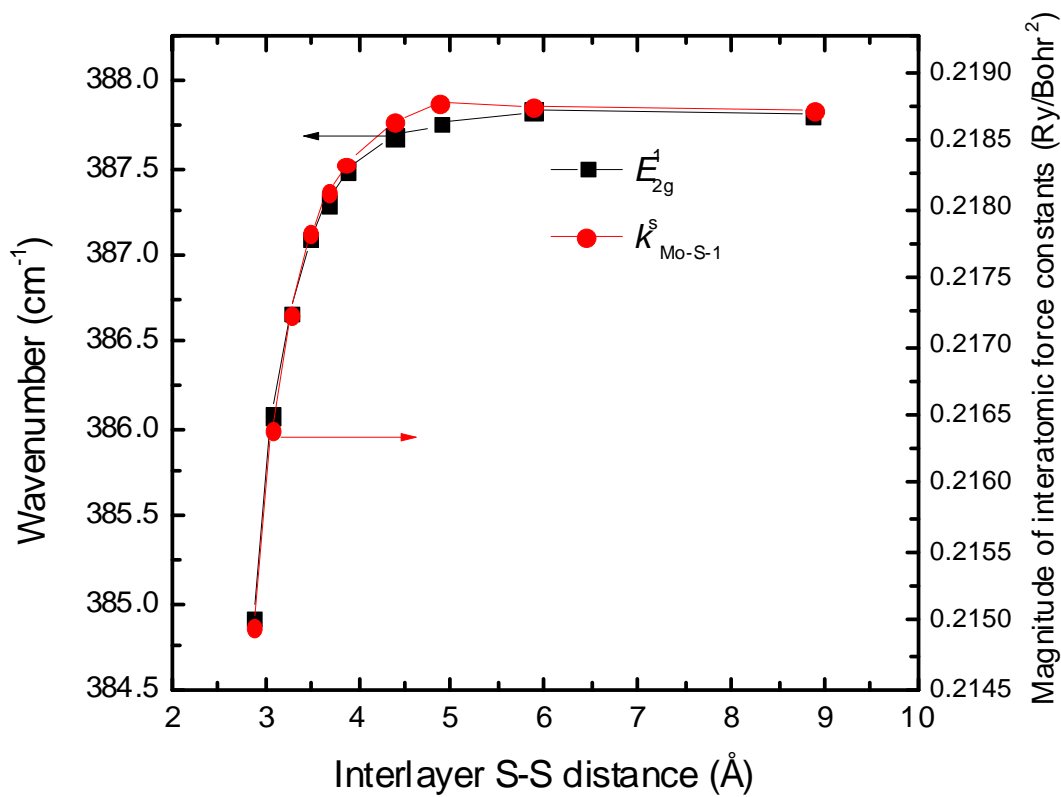

**Figure S5. The frequency of $E^1_{2g}$ mode and magnitude of $k^S_{Mo\text{-}S\text{-}1}$ as a function of the interlayer S-S distance in MoS$_2$ bulk structures.**

The equilibrium S-S interlayer distance in bulk MoS$_2$ is 2.87Å. With increasing interlayer S-S distance, both the frequency of $E^1_{2g}$ mode and the magnitude of $k^S_{Mo\text{-}S\text{-}1}$ increase rapidly and reach a stable value when the distance is larger than 4.37Å. Interestingly, the stable values of frequency and force constant are the same as those of the artificial 1TL system, in which the inter-atomic bond lengths are fixed to the optimized bulk structure (Fig. S4).